\documentclass{article}
\usepackage{arxiv}
\usepackage[utf8]{inputenc} 
\usepackage[T1]{fontenc}    
\usepackage{hyperref}       
\usepackage{url}            
\usepackage{booktabs}       
\usepackage{amsfonts}       
\usepackage{nicefrac}       
\usepackage{amsmath,amssymb,amsfonts}
\usepackage{cleveref}       
\usepackage{lipsum}         
\usepackage{graphicx}
\usepackage{cite}
\usepackage{doi}
\usepackage{algorithmic} 
\usepackage{textcomp}
\usepackage{multirow}

\title{BabelBrain: An Open-Source Application for Prospective Modeling of Transcranial Focused Ultrasound for Neuromodulation Applications.}

\date{January 29, 2023}

\author{ \href{https://orcid.org/0000-0002-7919-8587}{\includegraphics[scale=0.06]{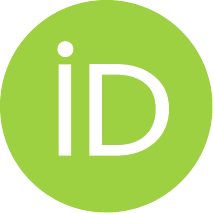}\hspace{1mm}Samuel Pichardo}\\
Department of Radiology and\\
Department of Clinical Neurosciences\\
University of Calgary\\
Calgary, AB, T2N 1N4\\
 Canada\\
 	\texttt{e-mail: samuel.pichardo@ucalgary.ca}\\
}

\renewcommand{\shorttitle}{BabelBrain: An Open-Source Application for Modeling of Transcranial Focused Ultrasound}

\hypersetup{
pdftitle={BabelBrain: An Open-Source Application for Prospective Modeling of Transcranial Focused Ultrasound for Neuromodulation Applications.},
pdfauthor={Samuel Pichardo},
pdfkeywords={transcranial ultrasound, modeling, open source, neuromodulation, neuronavigation.},
}

\begin{document}
\maketitle

\begin{abstract}
  BabelBrain is an open-source standalone graphic-user-interface application designed for studies of neuromodulation using transcranial focused ultrasound. It calculates the transmitted acoustic field in the brain tissue, taking into account the distortion effects caused by the skull barrier. The simulation is prepared using scans from magnetic resonance imaging (MRI) and, if available, computed tomography and zero-echo time MRI scans. It also calculates the thermal effects based on a given ultrasound regime, such as the total duration of exposure, the duty cycle, and acoustic intensity. The tool is designed to work in tandem with neuronavigation and visualization software, such as 3DSlicer. It uses image processing to prepare domains for ultrasound simulation and uses the BabelViscoFDTD library for transcranial modeling calculations. BabelBrain supports multiple GPU backends, including Metal, OpenCL, and CUDA, and works on all major operating systems including Linux, macOS, and Windows. This tool is particularly optimized for Apple ARM64 systems, which are common in brain imaging research. The paper presents the modeling pipeline used in BabelBrain and a numerical study where different methods of acoustic properties mapping were tested to select the best method that can reproduce the transcranial pressure transmission efficiency reported in the literature.
\end{abstract}

\keywords{transcranial ultrasound \and modeling \and open source \and neuromodulation \and neuronavigation}

\section{Introduction}
\label{sec:introduction}
\begin{figure}[!t]
  \centerline{\includegraphics[width=0.4\columnwidth]{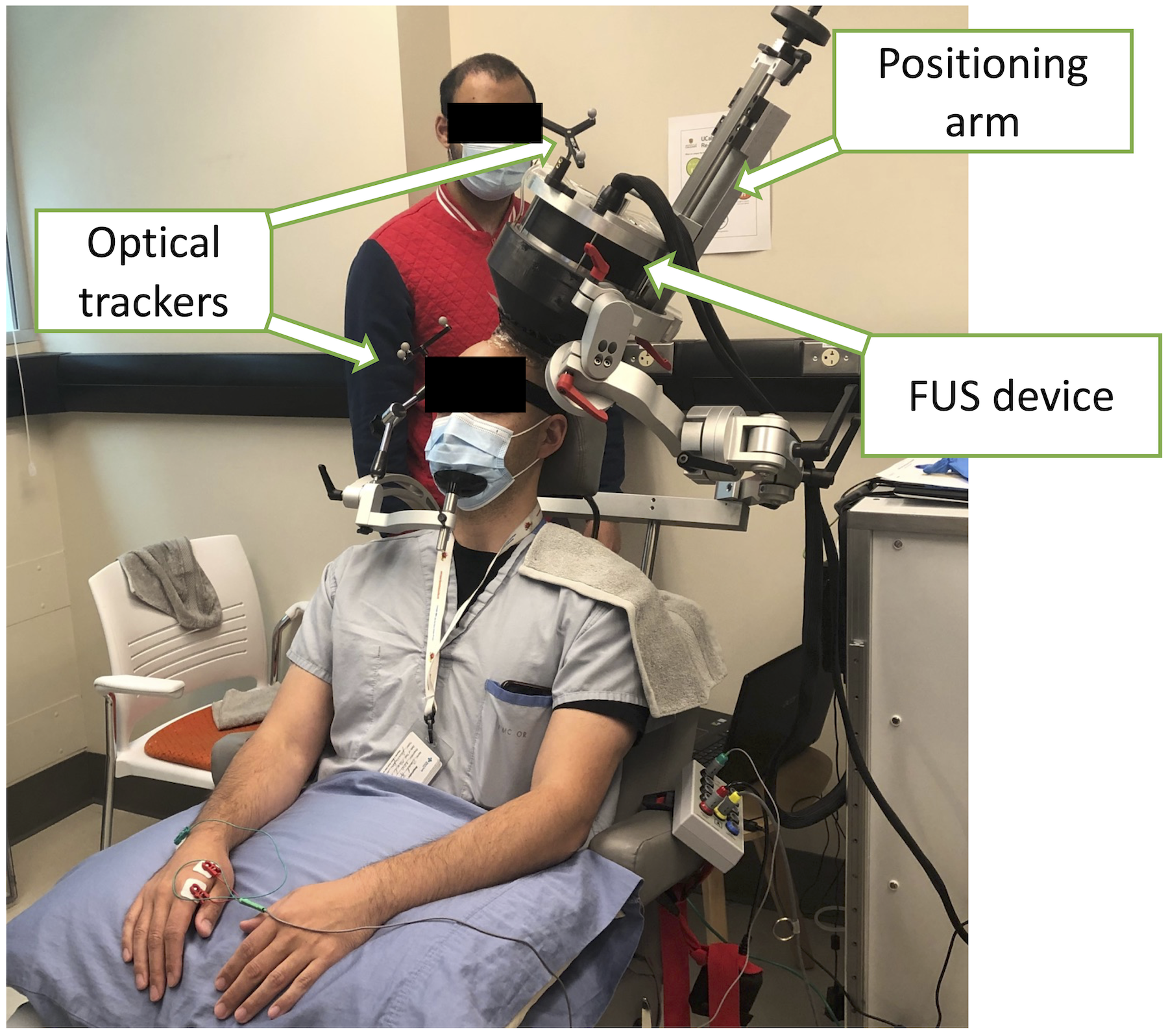}}
  \caption{Example of neuronavigated FUS device for neurostimulation experiments in humans at the Cumming School of Medicine of the University of Calgary. }
  \label{figNeuro}
  \end{figure}

Trasncranial focused ultrasound (FUS) applications for brain therapy have been studied since the early 20$^{\text{th}}$ century\cite{obrienEarlyHistoryHighintensity2015}. 
Combining high-intensity FUS to produce localized thermal ablation effects with Magnetic Resonance Imaging (MRI) for guidance and monitoring has led to FDA-approved noninvasive procedures such as treating essential\cite{eliasRandomizedTrialFocused2016c} and Parkinsonian\cite{bondSafetyEfficacyFocused2017} tremors. The combination of low-intensity FUS with microbubbles contrast agents has translated into a new type of cavitation-mediated therapy for brain indications\cite{mengSafetyEfficacyFocused2019,buneviciusFocusedUltrasoundStrategies2020,mengApplicationsFocusedUltrasound2021}. 
Recently, research has expanded on the non-lesioning neuromodulatory effects of FUS in the central nervous system (CNS). These effects, which can be traced back to the pioneering work of the Fry brothers in the 1950s\cite{barnardEffectsHighIntensity1955a}, have gained significant momentum in the last 10-15 years\cite{pasquinelliSafetyTranscranialFocused2019,blackmoreUltrasoundNeuromodulationReview2019}.
As shown in Fig. \ref{figNeuro}, experiments on humans with FUS-based 
neuromodulation\cite{leeImageGuidedTranscranialFocused2015,leeSimultaneousAcousticStimulation2016b,
legonTranscranialFocusedUltrasound2018a,
brinkerFocusedUltrasoundPlatform2020,
fomenkoSystematicExaminationLowintensity2020,
braunTranscranialUltrasoundStimulation2020,
yuTranscranialFocusedUltrasound2021,
leePilotStudyFocused2022,
brinkerFeasibilityUpperCranial2022,
zengInductionHumanMotor2022} are often conducted in setups analogous to other non-invasive neuromodulation techniques, such as transcranial magnetic stimulation, where neuronavigation systems assist in the targeting. These systems use optical tracking and preoperatory imaging of participants to position an ultrasound transducer relative to their target. However, the prediction of changes in ultrasound focusing and loss of energy level caused by the skull is not yet included prospectively in most neuronavigated procedures. 
As studies in humans of FUS-based neuromodulation multiply, there is a growing need to develop planning tools to help experimenters predict targeting and energy levels correctly, which are necessary to ensure the safety and harmonization of FUS delivery across participants.
Modeling transcranial ultrasound has been an intense research topic for the last 20 years.  Open-source and open-available tools such as 
SimSONIC \cite{bossyThreedimensionalSimulationsUltrasonic2004},
k-Wave \cite{treebyKWaveMATLABToolbox2010a},
FOCUS \cite{zhaoSimulationsNonlinearContinuous2017}, 
BabelViscoFDTD \cite{pichardoViscoelasticModelPrediction2017b}, 
Kranion \cite{sammartinoKranionOpensourceEnvironment2019},
HIFU Beam \cite{yuldashevHIFUBeamSimulator2021},
mSOUND \cite{guMSOUNDOpenSource2021}, 
J-Wave \cite{stanziolaJWaveOpensourceDifferentiable2022},
and Stride \cite{cuetoStrideFlexibleSoftware2022} (just to mention a few) have been used in multiple studies of characterization of transcranial ultrasound for therapy and imaging.  
Angla \emph{et al.} recently presented a thorough and excellent review on efforts of modeling of transcranial ultrasound for therapy\cite{angla_transcranial_2022}.
Several of the modeling tools mentioned above (along with proprietary software tools) have been cross-validated in simplified numerical settings of transcranial ultrasound\cite{aubry_benchmark_2022}. However, to the best of our knowledge, there are few to none open-accessible complete modeling suites that work in integration with neuronavigation systems to assist the planning of FUS-based neuromodulation experiments.
 
BabelBrain\footnote{\href{https://github.com/ProteusMRIgHIFU/BabelBrain}{https://github.com/ProteusMRIgHIFU/BabelBrain}} is an open-source application designed to assist in the prospective planning of FUS-based neurostimulation. It uses advanced modeling on transcranial ultrasound\cite{pichardoViscoelasticModelPrediction2017b,drainvilleSuperpositionMethodModelling2019}, incorporates the latest research on mapping acoustic properties, and is designed to be as accessible as possible. This paper presents the operational aspects of the tool and a numerical study on selecting the method for mapping acoustic properties. Studies on computing performance and the influence of the type of tissue mask generation method are also provided.

\section{Tool description and Modeling pipeline}
BabelBrain is a graphic user interface (GUI) application written in Python language, distributed as source code via GitHub and as a standalone application, eliminating the need for users to create specific Python environments. It is optimized to use modern GPU processors from Nvidia, AMD, and Apple to accelerate calculations and is available on all major operating systems, with a focus on ARM64 processors running on macOS. The software uses 3D MRI and, if available, computed tomography (CT) imaging data and a FUS trajectory. The latter represents a vector that indicates the transducer's focus position and acoustic axis alignment in T1w space. In practice, as shown in Fig. \ref{figPlanning}, the trajectory is a linear transformation applied to a virtual needle representing the focus location and acoustic axis. BabelBrain uses several libraries for GPU acceleration (CuPy, PyOpenCL, and metal-compute) depending on the available GPU and operating system. Currently, the following four types of transducers are supported:  
\paragraph*{Single} Simple focusing single-element transducer for which 
  the user can specify the diameter, focal length, and frequency between
  200 kHz and 1000 kHz\footnote{Frequencies larger than 800 kHz may require a computer system with 64 GB of RAM or more.}.
  \paragraph*{H317} 128-element phased array (SonicConcepts, Bothell, WA) with a focal length of 135 mm and F\#=0.9 capable of operating at 250 kHz and 700 kHz.
  \paragraph*{CTX-500} Four-element ring device 
  (NeuroFUS, Bothell, WA) with a focal length of 63.2 mm and F\#=0.98, and operating at 500 kHz.
  \paragraph*{H246} Flat ring-type device (NeuroFUS) with 2 ring
  elements, a diameter of 33.6 mm, and operating at 500 kHz. 
  \begin{figure}[!t]
    \centerline{\includegraphics[width=0.5\columnwidth]{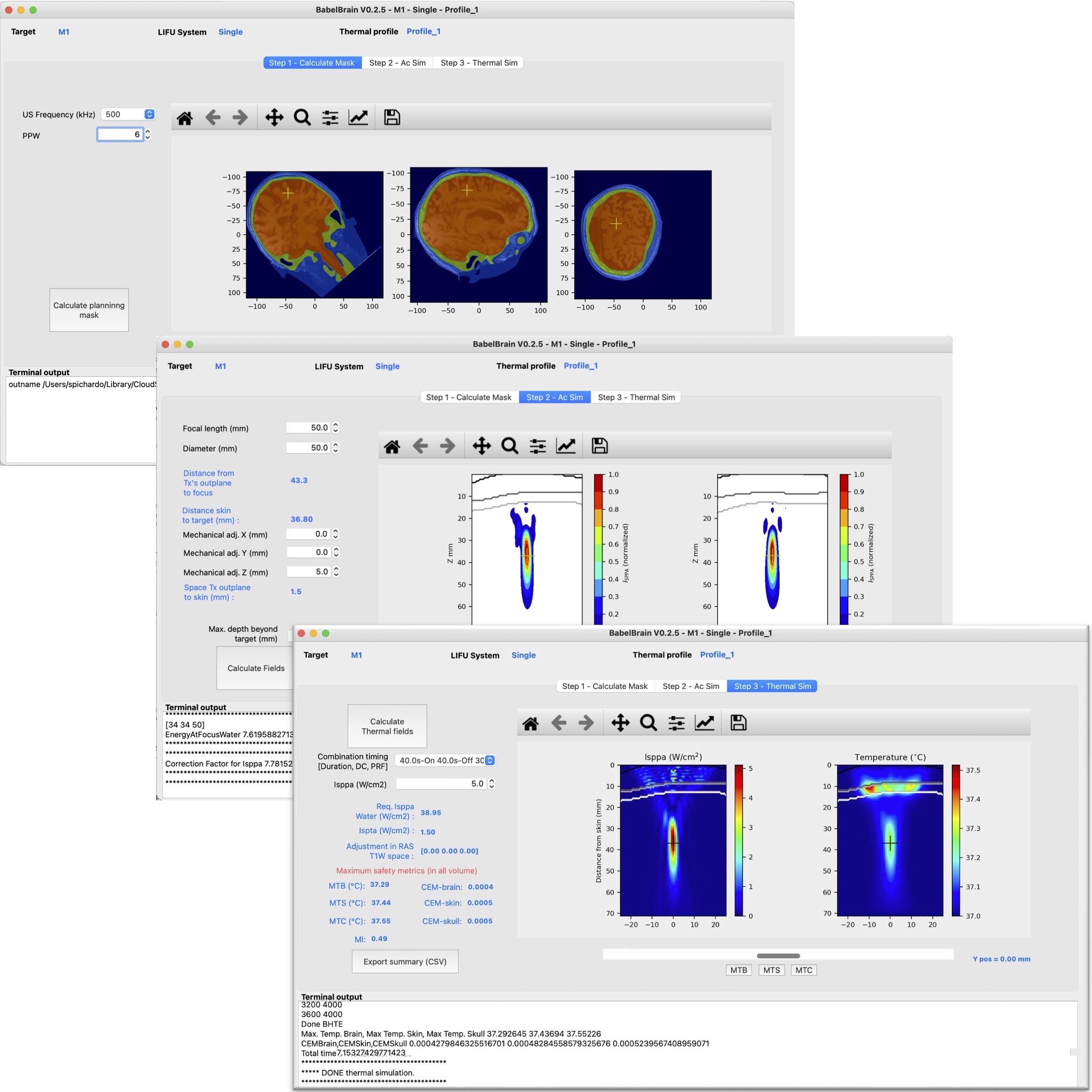}}
    \caption{BabelBrain's GUI execution steps. Top) Domain generation. Middle) Transcranial ultrasound modeling. Botton) Thermal modeling.}
    \label{FigGUI}
    \end{figure} 
Before modeling with BabelBrain, three preliminary steps need to be performed: imaging of the participant, preprocessing the subject's MRI data, and planning with neuronavigation or visualization software. As shown in Fig. \ref{FigGUI}, BabelBrain's operation comprises three sequential steps:  Domain generation, transcranial ultrasound modeling, and thermal modeling.

\subsection{Preliminary steps}\label{preliminary-steps}
As inputs, BabelBrain uses a text file describing the focused ultrasound trajectory, NifTI files of the participant's 3D isotropic 1mm resolution T1-weighted (T1w) scan, and NifTI files of segmented tissue produced with the help of the SimNIBS package \cite{thielscher_field_2015}, which is separate software for the simulation of transcranial magnetic and electric stimulation. T2-weighted (T2w) imaging is highly recommended for better tissue segmentation. If available, a NifTI file of a CT scan can improve the precision of simulations. Alternatively, if available, a NifTI file of a Zero Echo Time (ZTE) MRI scan can be used by BabelBrain to reconstruct a pseudo-CT using the so-called classical approach described by  Miscouridou \emph{et al.}\cite{miscouridouClassicalLearnedMR2022}. The segmented tissue masks are obtained using SimNIBS 3.x \texttt{headreco} or SimNIBS 4.x \texttt{charm} processing tools, which must be run before BabelBrain. 
\begin{figure}[!t]  
  \centerline{\includegraphics[width=0.45\columnwidth]{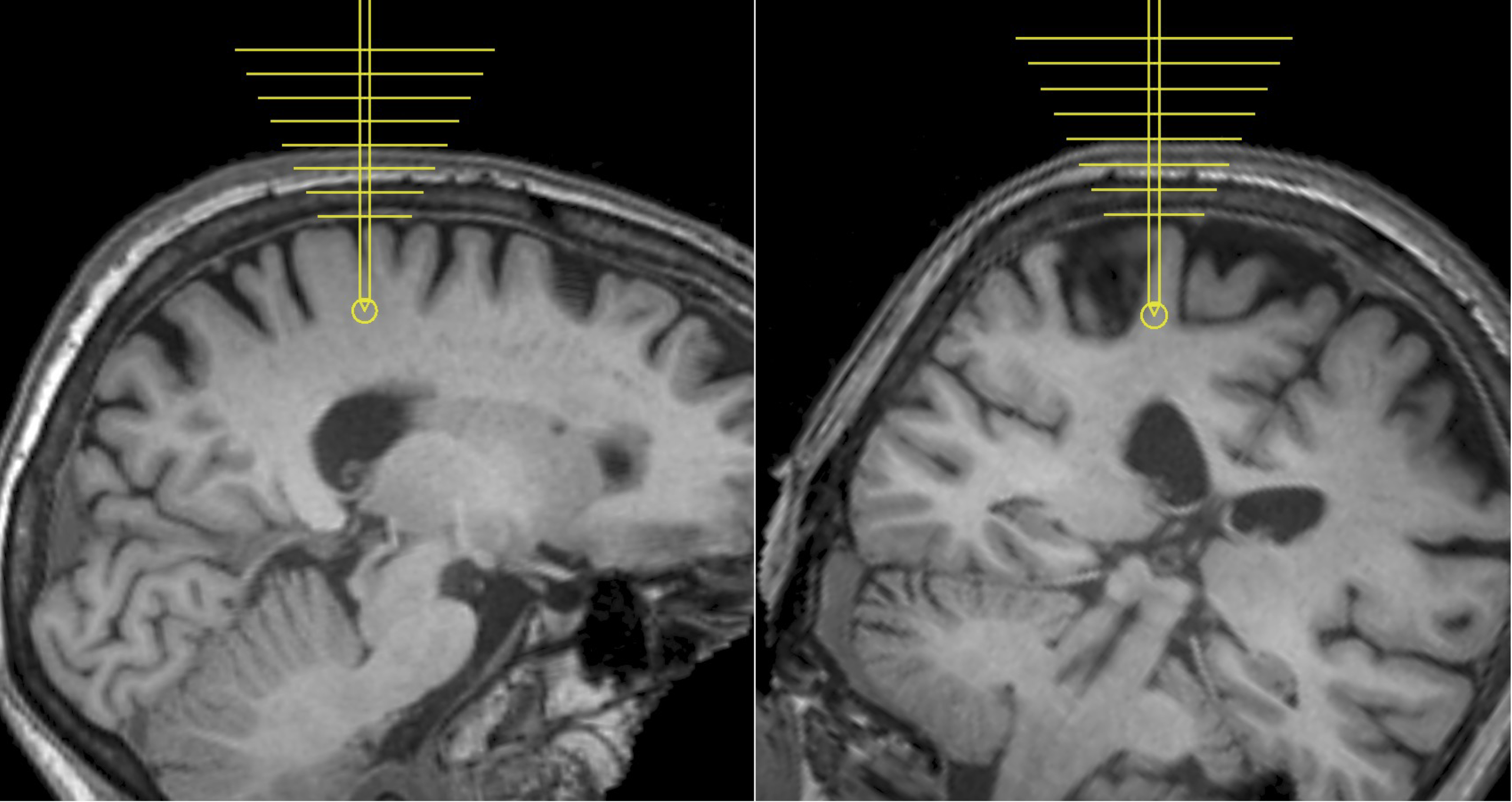}}
  \caption{Planning trajectory with 3DSlicer. A linear transform is applied to an STL object (yellow) provided with BabelBrain to mimic acoustic incident planes with an F\#=1 and align the trajectory to the intended target. The linear transform is connected to a re-slicer module (IGTSlicer) to generate in-plane (left) and rotated 90\textdegree~ (right) in-plane views.  }
  \label{figPlanning}
  \end{figure}
The planning trajectory file is a 4$\times$4 linear transformation matrix generated with the open-source visualization software 3DSlicer\cite{fedorov3DSlicerImage2012} or a proprietary neuronavigation software such as Brainsight (Rogue Research, Montreal, Canada).
Figure \ref{figPlanning} shows an example of the planning performed in 3DSlicer, where a linear transform and an object representing the acoustic path are used to create the trajectory.  Step-by-step instructions are detailed on the online documentation\footnote{\href{https://proteusmrighifu.github.io/BabelBrain}{https://proteusmrighifu.github.io/BabelBrain}}.

\subsection{Domain generation}\label{domain-generation}
Different processing is executed depending on the available imaging inputs. The spatial resolution, based on frequency and points-per-wavelength (PPW) in water, is selected in this step.
\subsubsection{Simplified mask}  
A label mask (skin, skull, and brain) with a resolution based on the frequency and PPW is produced using the preprocessed output data via \texttt{headreco} or \texttt{charm}. The preprocessed outputs are used to generate a Gaussian-filtered isosurface, voxelized, and median-filtered at a higher resolution (0.75 PPW). The raw 3D label mask is oriented with the $z$ direction aligned to the ultrasound transducer's acoustic axis. The data is saved in NifTIi format to preserve spatial registration in T1w space. The skull bone is split into 3 layers, a medium of cortical (10\%), trabecular (80\%), and cortical (10\%) distribution. Table  \ref{tableMaterial} shows material properties for homogenized conditions.
\begin{table*}
  \caption{Material properties for homogenized materials. $c_L$: longitudinal speed. $c_S$: Shear speed. $\alpha_L$: longitudinal attenuation. $\alpha_S$: Shear attenuation. $f$: Frequency (MHz). $^{\triangle}$ \cite{duckPhysicalPropertiesTissues2013}. $\star$ Linear interpolation from \cite{pichardo_multi-frequency_2010}. $\dagger$  Linear interpolation from \cite{pichardoViscoelasticModelPrediction2017b}. $\ddagger$ Average from \cite{gossComprehensiveCompilationEmpirical1978,pichardo_multi-frequency_2010,webb_acoustic_2021}. $\diamond$ Average from \cite{pichardoViscoelasticModelPrediction2017b}.}
  \begin{center}
  \begin{tabular}{lrrrrr}
    \toprule
    Material/ &  $\rho$   &  $c_L$   &  $c_S$  &   $\alpha_L$ &  $\alpha_S$  \\
    tissue  &     &  (m/s)&    (m/s)   &   (Np/m/MHz)  &   (Np/m/MHz)   \\
    \midrule
    Water &              1000   & 1500  &   - &  - & -  \\
    Skin &               1116$^{\triangle}$   & 1537$^{\triangle}$   &  -  & 4.6$f^1$$^{\triangle}$   &  -  \\
    Brain &              1041$^{\triangle}$   & 1562$^{\triangle}$   &  -  & 6.9$f^1$$^{\triangle}$  & -  \\
    Cortical bone &      1896.5 & 2415+$1.2$e-4$f$$\star$ & 1369+$3.4$e-4$f$$\dagger$ & 163$f^1$$\ddagger$ & 329$f^1$$\diamond$  \\
    Trabecular bone &    1738.0 & 2063+$2.8$e-4$f$$\star$ & 1206+$2.1$e-4$f$$\dagger$  & 162$f^1$$\ddagger$ &  329$f^1$$\diamond$ \\
    \bottomrule
    \end{tabular}
  \end{center}
  \label{tableMaterial}
  \end{table*}

\subsubsection{CT-based mask} \label{subctbased}
When a CT scan is available, the label masks are refined to include voxel-level information of the skull bone. The T1w scan is corrected for bias\cite{tustisonN4ITKNickN32010} and upscaled to match the CT scan resolution, then co-registered to T1w space using the elastix software v5.0.1\cite{kleinElastixToolboxIntensityBased2010} (included in BabelBrain). A user-defined threshold (default to 300 HU) is applied, and a refined mask of skull bone is obtained by producing a Gaussian-filtered isosurface, voxelized, and median-filtered at a higher resolution (0.75 PPW). CT values in the masked region are quantified at discrete values with 12 bits of resolution. The mapping from HU to density, speed of sound, and attenuation is detailed in the numerical study presented in section \ref{mappingsection}.

\subsubsection{ZTE-based mask generation} 
ZTE scans can be used to generate synthetic CT scans using a  classical method\cite{miscouridouClassicalLearnedMR2022}. This method was developed for applications of radiation attenuation correction in Positron Emission Tomography with MRI\cite{wiesingerZeroTEbasedPseudoCT2018} and used a normalization of the ZTE bone signal using the soft tissue ZTE signal. A user-specified range of normalized ZTE values (with defaults of 0.1 and 0.6) determines the region identified as skull bone. The conversion to synthetic CT maps is done as presented in Miscouridou \emph{et al.}\cite{miscouridouClassicalLearnedMR2022}, followed by the processing for CT scans mentioned in \ref{subctbased}.

\subsection{Transcranial ultrasound modeling}\label{transcranial-modeling}
Transcranial ultrasound modeling is performed using the open-source BabelViscoFDTD\footnote{\href{https://github.com/ProteusMRIgHIFU/BabelViscoFDTD}{https://github.com/ProteusMRIgHIFU/BabelViscoFDTD}} library, which uses a GPU-accelerated finite-difference time-difference (FDTD) solver for the viscoelastic equation in isotropic conditions\cite{pichardoViscoelasticModelPrediction2017b,drainvilleSuperpositionMethodModelling2019} given by\begin{footnotesize}
  \begin{equation}
  \begin{aligned}
  \dfrac{\partial v_x}{\partial t} & =  \dfrac{1}{\rho} \left( \dfrac{\partial \sigma_{xx}}{\partial_x} 
  + \dfrac{\partial \sigma_{xy}}{\partial_y} + \dfrac{\partial \sigma_{xz}}{\partial_z} \right),   \\
  \dfrac{\partial v_y}{\partial t} & =  \dfrac{1}{\rho} \left( \dfrac{\partial \sigma_{xy}}{\partial_x} 
  + \dfrac{\partial \sigma_{yy}}{\partial_y}  + \dfrac{\partial \sigma_{yz}}{\partial_z}\right), \\
  \dfrac{\partial v_z}{\partial t} & =  \dfrac{1}{\rho} \left( \dfrac{\partial \sigma_{xz}}{\partial_x} 
  + \dfrac{\partial \sigma_{yz}}{\partial_y}  + \dfrac{\partial \sigma_{zz}}{\partial_z}\right), \\
  \dfrac{\partial \sigma_{xx}}{\partial t} & =  (\Lambda + 2 M) \dfrac{\partial v_x}{\partial_x} + \Lambda
   \left( \dfrac{\partial v_y}{\partial y} +\dfrac{\partial v_z}{\partial z} \right), \\
  \dfrac{\partial \sigma_{yy}}{\partial t} & =  (\Lambda + 2 M) \dfrac{\partial v_y}
  {\partial_y} + \Lambda 
  \left( \dfrac{\partial v_x}{\partial x} + \dfrac{\partial v_z}{\partial z} \right), \\
  \dfrac{\partial \sigma_{zz}}{\partial t} & =  (\Lambda + 2 M) \dfrac{\partial v_z}
  {\partial_z} + \Lambda 
  \left( \dfrac{\partial v_x}{\partial x} + \dfrac{\partial v_y}{\partial y} \right), \\
  \dfrac{\partial \sigma_{xy}}{\partial t} & =  M \left( \dfrac{\partial v_x}{\partial_y} + \dfrac{\partial v_y}{\partial_x}   \right)\\
  \dfrac{\partial \sigma_{xz}}{\partial t} & =  M \left( \dfrac{\partial v_x}{\partial_z} + \dfrac{\partial v_z}{\partial_x}   \right)\\
  \dfrac{\partial \sigma_{yz}}{\partial t} & =  M \left( \dfrac{\partial v_y}{	\partial_z} + \dfrac{\partial v_z}{\partial_y}   \right)
  \end{aligned}
  \label{Eq:Viscoelastic}
  \end{equation}
  \end{footnotesize} 
  where $v_i$ and $\sigma_{ij}$ are the displacement vectors and stress tensors, respectively, in each of $i,j$-directions: $x$, $y$ or $z$. $\Lambda$ and $M$ are the Lam\'e parameters of the material. As shown in previous work \cite{pichardoViscoelasticModelPrediction2017b}, the equation system \ref{Eq:Viscoelastic} allows modeling of both linear compressional and shear wave transmission, including mode conversion between water and bone materials. The solver uses a 4th-order in space and 2nd-order in time FDTD scheme in Cartesian coordinates \cite{levanderFourthOrderFinite1988,bohlenParallel3DViscoelastic2002} with a staggered-grid arrangement \cite{virieuxPSVWavePropagation1986}. 
   Details of implementation, including attenuation modeling, temporal step calculation, and boundary matching conditions, can be consulted elsewhere \cite{pichardoViscoelasticModelPrediction2017b}.

BabelBrain's simulation is customized to the type of transducer selected. For example, electronic steering in depth for the CTX500 and steering in all three directions, and refocusing for the H317 transducer. For a single transducer, the user can specify the aperture, focal length, and distance to the skin. The selected transducer is modeled by numerical surface decomposition with surface elements of an area of $\frac{\lambda^2}{16}$, where $\lambda$ is the wavelength in water. The acoustic field is first calculated with a GPU-accelerated implementation of the Rayleigh-Sommerfeld equation\cite{pierceAcousticsIntroductionPhysical1994} over the incident interface of the simulation domain, with the coupling medium modeled as water. The incident field is then forward-propagated with the FDTD solver\cite{pichardoViscoelasticModelPrediction2017b} as a stress source perpendicular to the acoustic axis. A water-only field is also calculated, later used to calculate the thermal effects. For the H317 transducer, if refocusing is requested, a punctual source is placed at the intended target, back-propagated to the domain interface, and then propagated back with Rayleigh-Sommerfeld to the transducer elements to calculate the corrected phase programming of the transducer elements. Finally, the forward propagated field is calculated.

\subsection{Thermal modeling}
The thermal simulation step solves the Bio Heat Transfer Equation (BHTE) with a finite-difference time-difference solver in Cartesian coordinates\cite{pichardoCircumferentialLesionFormation2007c}. Boundary conditions are assumed conservatively with a temperature of 37\textdegree C at the domain's borders. The user specifies the sonication regime to be studied with the duration of ultrasound, time off after ultrasound, and duty cycle. The user also indicates the desired spatial-peak pulse-average acoustic intensity ($I_{\text{SPPA-brain}}$) in brain conditions after considering all tissue losses. The fraction of attenuation in skull bone transformed into heat was set to 0.16 based on findings by Pinton \emph{et al.}\cite{pintonAttenuationScatteringAbsorption2012}. Safety metrics are calculated as a function of parameter selection, including the mechanical index in the brain region, maximal temperature, and thermal dose\cite{damianouEvaluationAccuracyTheoretical1995} in the brain, skin, and skull. 

The thermal modeling step also calculates the required intensity in water conditions ($I_{\text{SPPA-water}}$) that would produce a requested $I_{\text{SPPA-brain}}$. It is assumed that the experimenters will have access to calibration information indicating $I_{\text{SPPA-water}}$ (or peak pressure) as a function of hardware programming (e.g., input voltage, electrical power, etc.). For example, the CTX-500 device allows experimenters to select $I_{\text{SPPA-water}}$ in the driving unit. An energy loss $E_L$ factor is calculated with
\begin{equation}
E_L=\frac{\sum_{x,y} I_{\text{PA-brain}}(x,y,z_m)}{\sum_{x,y}I_{\text{PA-water}}(x,y,z_m)},
\end{equation}
where $I_{\text{PA}}$ is the intensity peak average. 
 $E_L$ is calculated over the plane at a distance $z_m$ where water conditions show the maximal intensity and is used to calculate the required  $I_{\text{SPPA-water}}$ for a given  $I_{\text{SPPA-brain}}$. The results from the transcranial modeling step are initially scaled to match a default value $I_{\text{SPPA-brain}}=5$ W/cm$^2$. User's changes to $I_{\text{SPPA-brain}}$ recalculates safety metrics using 5 W/cm$^2$ as a baseline.

\section{Numerical investigation on mapping functions between CT scans and acoustic properties}
\label{mappingsection}
The complex composition of the skull bone causes significant distortion of the ultrasound transmission into the brain cavity. To date, there is no definitive method for producing maps of acoustic properties (density, speed of sound, and attenuation) from CT or MRI data. Different approaches to producing these maps have been reviewed by Angla \emph{et al}\cite{angla_transcranial_2022}. Acoustic maps derived from CT scans facilitated the development of procedures approved for use in clinic\cite{elias_randomized_2016}, but this process has been in a very specific set of conditions. Webb \emph{et al.} presented two studies detailing clearly how XRays energy level, scanner model, and CT kernel reconstruction influence acoustic properties derived from Hounsfield units (HU)\cite{webb_measurements_2018,webb_acoustic_2021}. A numerical study was performed to select the most appropriate mapping method to use with BabelBrain.

\subsection{Methodology}
A literature review was conducted to identify mapping methods that converted Hounsfield units (HU) to both the speed of sound and attenuation values. These methods were tested using five CT scans that represent a range of skull density ratios (SDR). The SDR is a metric that measures the HU of trabecular bone relative to cortical bone, and is used in treating essential and Parkinsonian tremors with MRI-guided focused ultrasound (MRIgFUS). Values for SDR typically range between 0.3 and 0.8 and have been identified as one of the factors determining the effectiveness of producing FUS thermal effects for surgical purposes\cite{chang2016factors,boutet2019relevance}. The lower the SDR, the less effective the MRIgFUS procedure is in inducing thermal rising at the surgical target. The SDR is, therefore, an indirect metric associated with the global insertion losses of the skull bone. Based on the results of these tests, the mapping procedure that predicted pressure transmission losses closest to reported values in the literature was chosen for standard operation in the BabelBrain tool. A literature review was done to identify characterization studies that performed experimental data collection of pressure or energy transmission with frequencies less or equal than 1 MHz. 

\subsection{Methods}
Transcranial ultrasound simulations were performed with CT scans obtained for patients treated for essential tremor and obsessive-compulsive disorder with MRIgFUS at the University of Calgary. A study was approved by the University of Calgary Ethics Board to select five subjects representing a
sampling of SDR values to be available as part of this study. Independently of the original cohort of patients, SDR values were calculated by selecting the left ventral intermediate nucleus in the thalamus. Datasets are available at a Zenodo repository\footnote{URL and DOI to be available at publication time}. SDR was calculated by the planning software of
the MRIgFUS Neuro system (Exablate, Haifa, Israel). The selected SDR
values were 0.32, 0.44, 0.55, 0.67, and 0.79. 
\subsubsection{Imaging sourcing} CT imaging was performed with a
Discovery CT750 HD scanner (GE, MW, USA) with an energy level of 120
kVp, BONEPLUS reconstruction kernel, pixel resolution of 0.45 mm, slice
thickness of 0.625 mm and space between slices of 0.66 mm. The skull bone region was thresholded between 300 and 2100 HU. 
T1w and T2w scans were also acquired using a Discovery 750 3T MRI scan (GE, MW, USA). T1w imaging was
acquired using BRAVO, GE's 3D inversion recovery-prepared
fast spoiled gradient echo sequence with the following
parameters:
TR/TE/TI=8.2/3.2/650ms, 10\textdegree~ flip angle, 256$\times$256$\times$188mm field of view, and 256$\times$256$\times$188 acquisition matrix.
T2w imaging was acquired using 3D fast spin-echo with the
following parameters: TR/TE=3000,60--90ms, echo train
length=130, 256$\times$225$\times$188mm field of view, and  256$\times$256$\times$188 acquisition matrix (zero-filled to 512$\times$512$\times$376). MRI scans were processed with
\href{https://github.com/simnibs/simnibs/releases}{SimNIBS's}
\texttt{charm} tool\cite{thielscher_field_2015}. 

\subsubsection{Mapping approaches}
The following mapping methods to produce acoustic properties from CT scans were evaluated.

\paragraph*{Aubry}
  This method calculated maps of density, attenuation, and speed of sound to an approximation of the porosity of skull bone\cite{aubryExperimentalDemonstrationNoninvasive2003b}. A small modification, as done in Liu \emph{et al.}\cite{liuEvaluationSyntheticallyGenerated2022}, to the method of porosity calculation is applied to use maximum HU in the skull bone instead of a fixed value. 
  \paragraph*{McDannold} This method was developed using analysis of the thermal rise in
  essential tremor procedures with MRIgFUS and transcranial ultrasound
  modeling \cite{mcdannold_elementwise_2019}. McDannold calculated
  fitted coefficients between apparent density derived from CT scans and
  speed of sound and attenuation. Attenuation fitting from McDannold was
  scaled linearly with a factor
  $\left({f/0.66~\text{MHz}}\right)^1$ where $f$ is the
  frequency in MHz.
  \paragraph*{Pichardo} This method fitted the apparent density calculated from HU to the speed of sound and attenuation at different individual
  frequencies\cite{pichardo_multi-frequency_2010}. A linear
  interpolation was performed between the reported attenuation curves
  for frequencies between 0.27 and 0.836 MHz. For frequencies between
  0.1 and 0.27 MHz, the attenuation was scaled linearly with a factor
  $\left({f/0.27~\text{MHz}}\right)^1$. For frequencies
  between 0.836 and 1 MHz, a similar factor
  $\left({f/0.836~\text{MHz}}\right)^1$ was applied.
\paragraph*{Webb-Marsac} Webb presented methods to convert directly from HU to speed of
  sound\cite{webb_measurements_2018} and
  attenuation\cite{webb_acoustic_2021}. Because Webb's methods did not include mapping of density, the method by Marsac \emph{et al.}\cite{marsac_ex_2017} to convert from HU to density was used. Fitting parameters were
  selected for a GE scanner using an energy of 120 kVp, BONEPLUS kernel
  reconstruction, and axial and slice resolutions of 0.5 mm and 0.6 mm, respectively.

  Previous work\cite{pichardoViscoelasticModelPrediction2017b} established mapping of the shear speed of sound derived from CT scans. In that study, the Pichardo method for longitudinal properties was used to optimize the shear speed of sound. When using a viscoelastic solution, the longitudinal and shear speeds of sound are converted into Lam\'e coefficients. A preliminary test was performed to combine the functions for shear speed of sound from \cite{pichardoViscoelasticModelPrediction2017b} with the Aubry, McDannold, Webb-Marsac methods, but that translated that few voxels (around less than 0.1\% of skull region) produced Lam\'e coefficients that translated into Poison ratios slightly off the realistic range of values between [-1.0, 0.5]. Given the testing conditions in the present study are with a normal incidence of ultrasound to the skull interface, the mapping of shear speed properties from CT scans was disabled.

\subsubsection{Modeling}  
\begin{figure}[!t]
  \centerline{\includegraphics[width=0.5\columnwidth]{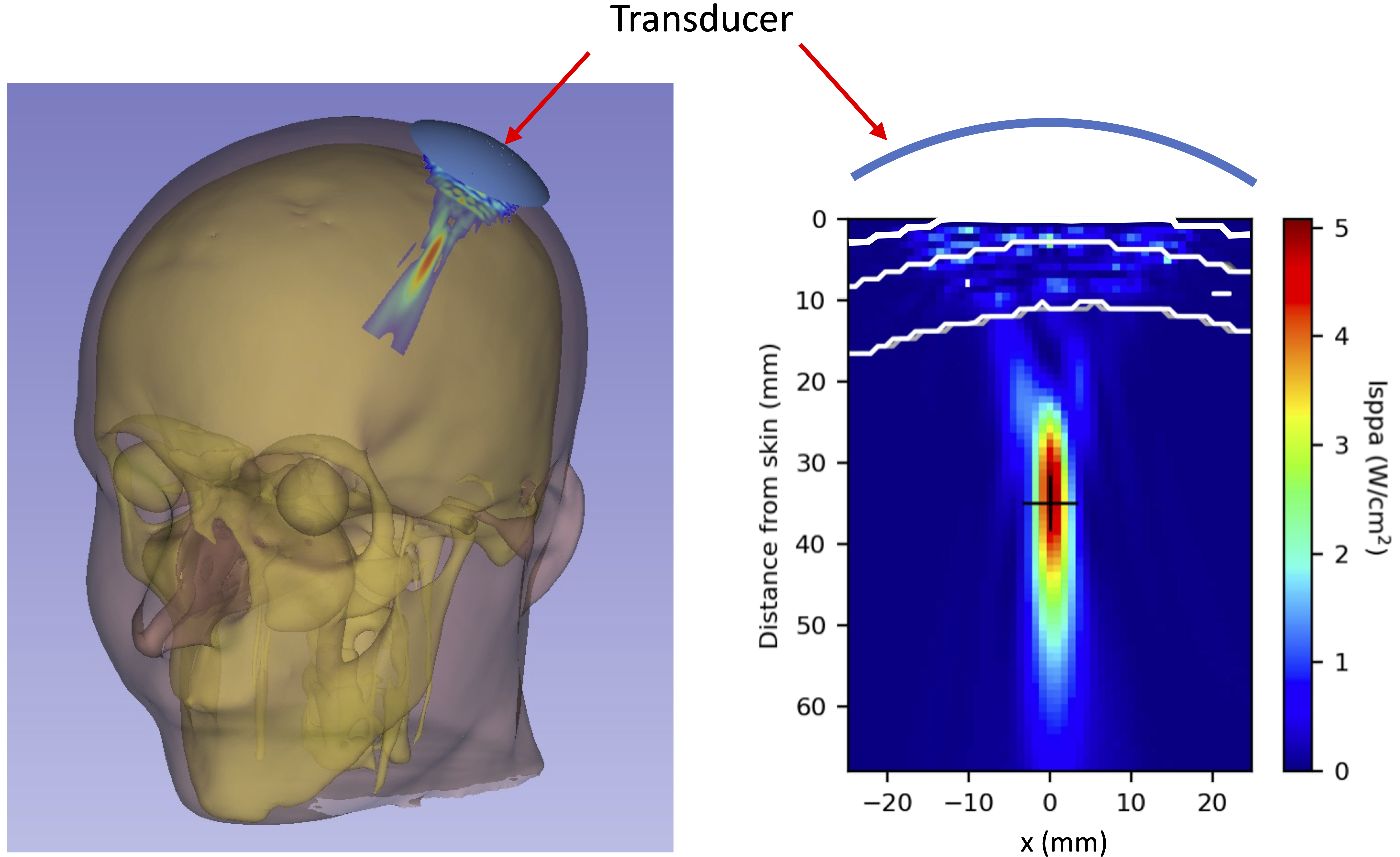}}
  \caption{Example of setup for simulations with the dataset of SDR=0.55 with pressure field at 600 kHz. The transducer was positioned 3 mm above the skin, with a trajectory with a normal incidence to the skin interface. 3D rendering of skin and bone surfaces, with a central cut of acoustic field along the acoustic axis (left). 2D detail of the acoustic field where the white lines show the tissue interfaces (right). The cross mark ($+$) indicates the intended target. }
  \label{figSetup}
  \end{figure}

Simulations were done for  each subject in the region below the M1 motor cortex 
35 mm below the skin. As shown in Fig. \ref{figSetup}, a transducer with a diameter
of 50 mm and F\# of 1.0 was placed 3 mm above the skin region with a
trajectory that ensured normal incidence conditions using 3DSlicer and the SlicerIGT module\cite{ungiOpensourcePlatformsNavigated2016}. Fig. \ref{figSetup} also shows an example of a central pressure map. The simulation domain had dimensions of 54$\times$54$\times$76 mm. Skin and brain regions were modeled as water to enable comparison with transmission pressure data in the literature.   

 Simulations were done for frequencies between 0.2 to 1
MHz, with a step of 0.1 MHz, with a spatial resolution of 6 PPW. Attenuation losses are modeled using a $Q$ factor\cite{alma991012368939704336},
which describes the loss of energy after traveling a $\pi$-distance,
as shown by

\begin{equation}
Q=\frac{\omega}{2c\alpha(\omega)},
\end{equation}

where $\omega$ is angular frequency, $c$ is the speed of sound and
$\alpha(\omega)$ the attenuation in Np/m at the frequency $\omega$.
In practice, the underlying numerical solver in
BabelViscoFDTD uses an optimization of relaxation constants and memory variables to match a given value of $Q$\cite{pichardo_multi-frequency_2010}.

A challenge when using a given mapping procedure is separating absorption and scattering, which can lead to large discrepancies in the results of mapping HU to attenuation\cite{webb_acoustic_2021}. Pinton \emph{et al.} demonstrated through viscoelastic and thermal modeling combined with micro CT scans of the skull bone that absorption is a small fraction of the total attenuation; 2.7 dB/cm at 1 MHz of pure absorption losses vs. 16.6 dB/cm of total attenuation\cite{pintonAttenuationScatteringAbsorption2012}. CT scans in clinical systems often have a resolution of 0.5 mm or larger, which can cause significant partial volume effects due to the microstructure of the bone material, especially in the trabecular region. Due to this ambiguity in the attenuation mapping, BabelViscoFDTD allows a correction factor $q_c$ to obtain $Q=Q_0q_c$, where $Q_0$ is the original associated with a given value of $\alpha(\omega)$. For each combination of values of SDR and frequency, transcranial modeling was done for values of $q_c$ of 0.1,0.2.,0.3,0.4,0.6, 0.8, 1.0, 1.5, 2.0, 3.0, 4.0, 5.0, 6.0, and 7.0.

\begin{figure}[!t]
  \centerline{\includegraphics[width=0.5\columnwidth]{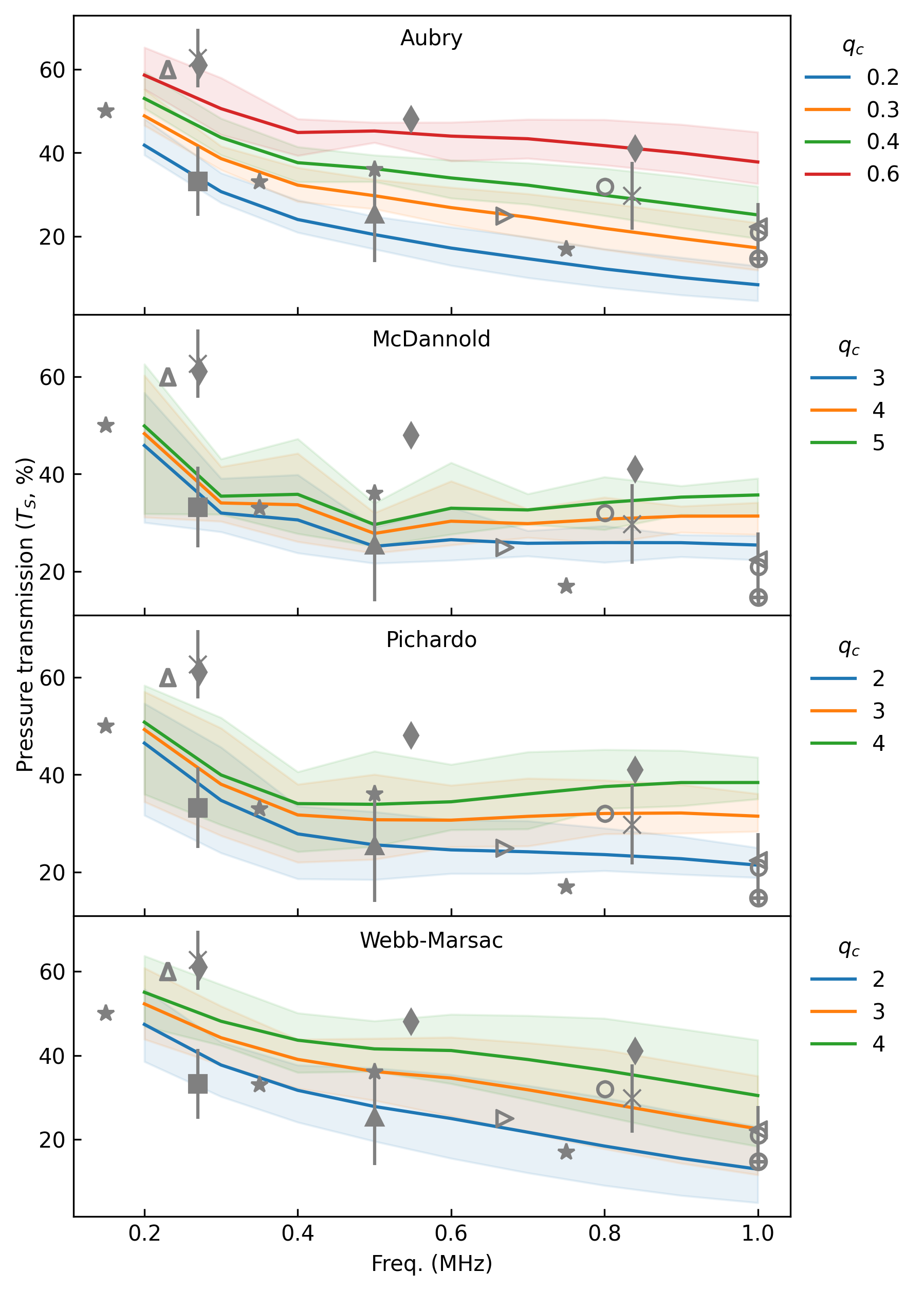}} 
  \caption{Effect of correction factor $q_c$ on transmitted pressure $T_S$. Only a subsample of $q_c$ values is shown for clarity purposes. The colored areas indicate the spread of prediction of $T_S$ across all datasets with different SDR values. $T_E$ values from the literature  are shown by 
  Alkins $\Delta$ \cite{alkinsCavitationbasedThirdVentriculostomy2013},
  Chen $\star$\cite{chenNumericalExperimentalEvaluation2023},
  Fry $\triangleleft$ \cite{fryTranskullTransmissionIntense1977b},
  Gimeno $\blacksquare$ \cite{gimenoExperimentalAssessmentSkull2019},
  Leung $\triangleright$ \cite{leungComparisonMRCT2022},
  Marsac $\bigcirc$\cite{marsac_ex_2017},
  Pichardo $\times$ \cite{pichardo_multi-frequency_2010},
  Pinton $\bigoplus $ \cite{pintonAttenuationScatteringAbsorption2012},
  Riis $\blacktriangle$\cite{riisAcousticPropertiesHuman2022},
  and White $\blacklozenge$\cite{whiteLongitudinalShearMode2006b}.
   Error bars indicate standard deviation.}
  \label{figQceffect}
  \end{figure}
  \begin{figure}[!t]  
    \centerline{\includegraphics[width=0.5\columnwidth]{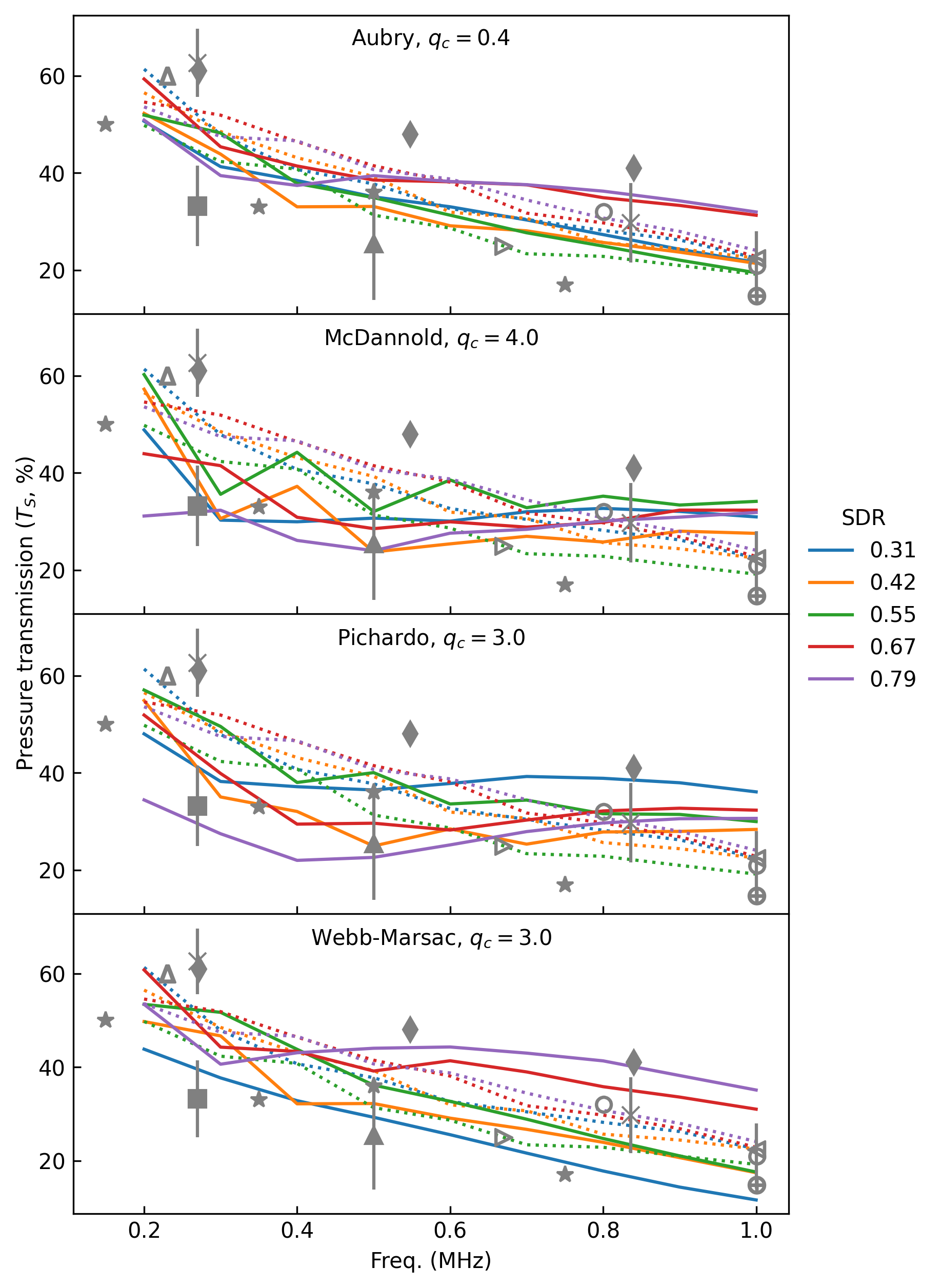}} 
    \caption{Detail of $T_S$ with mapping methods for each SDR value for transmitted pressure $T_S$ using its respective optimal correction factor $q_c$. $T_E$ values from literature are shown as in Fig. \ref{figQceffect}. Results with simplified masks obtained with only T1w and T2w input data are shown by the dotted lines ($\cdots$) plots.}
    \label{figDetailComparisonMap}
    \end{figure}

A pressure transmission coefficient $T_S$ was calculated with $T_S=\sqrt{E_L}$. The values of $T_S$ calculated with each of the different mapping approaches were compared against experimental values of sub-MHz pressure transmission ($T_E$) from literature reports on pressure \cite{pintonAttenuationScatteringAbsorption2012,
whiteLongitudinalShearMode2006b,
leungComparisonMRCT2022,
marsac_ex_2017,
gimenoExperimentalAssessmentSkull2019,
riisAcousticPropertiesHuman2022,
chenNumericalExperimentalEvaluation2023} and energy\cite{pichardo_multi-frequency_2010,
alkinsCavitationbasedThirdVentriculostomy2013,fryTranskullTransmissionIntense1977b} transmission. For the latter, $T_E$ was approximated by the square root of the reported energy transmission. Quality of mapping methods was calculated by the average distance of reported $T_E$ (expressed as a fraction from 0 to 1.0) to the polygonal area of values enclosed by $T_S(f,\text{SDR})$, where $f$ is expressed as a fraction of 1 MHz. The distance to this area was calculated with the function \texttt{geometry.Polygon.Distance} of the shapely v2.0 Python library. This function calculated a distance equal to 0 if $T_E$ is inside the area $T_S(f,\text{SDR})$. Otherwise, it calculates the Cartesian distance to the closest edge of the polygon.
 Simulations were executed in an iMac Pro computer system (Apple, Cupertino, CA) equipped with a 3 GHz Xeon-W
processor (10 cores), 128 GB RAM, and an external  W6800 Pro GPU (AMD, Santa Clara, CA) with 32 GB RAM and a MacBook Pro M1 Max with 64 GB RAM. 

\begin{figure*}
  \centerline{\includegraphics[width=1\columnwidth]{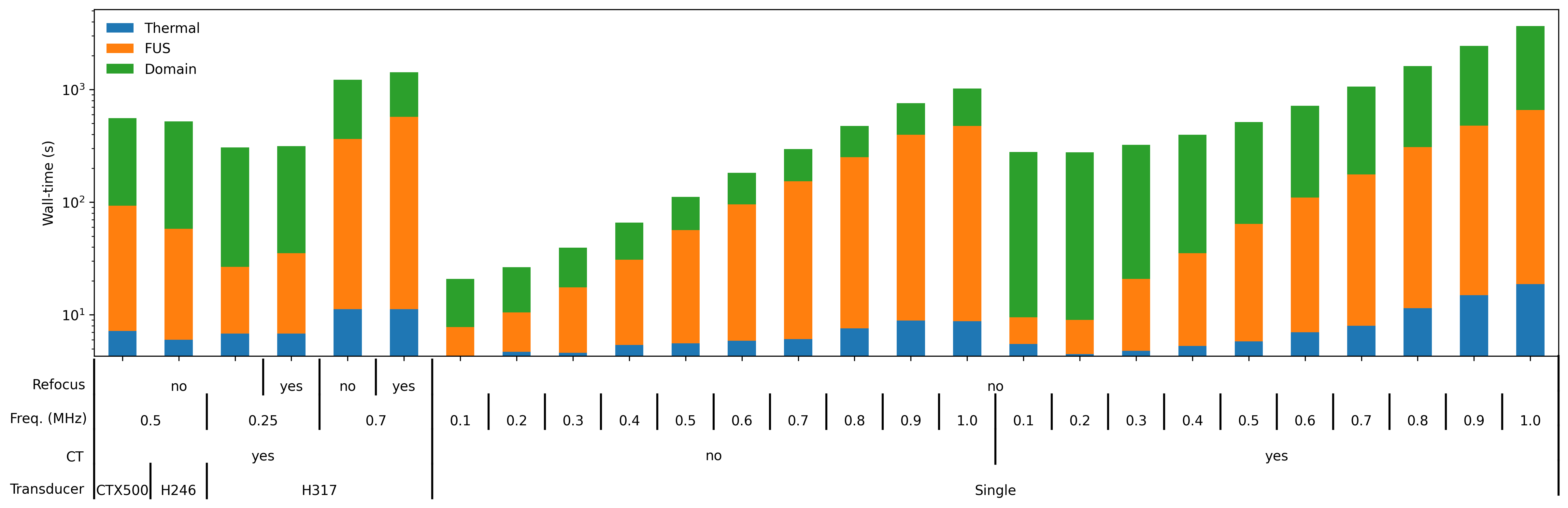}}
  \caption{Processing wall-time of multiple conditions of transducer, frequency, use of CT, and refocus (if available in a transducer) with a MacBook Pro M1 Max 64 GB. Wall-time is plotted in a logarithm scale and stacked from the procedure's fastest to the slowest step. Domain = Domain generation step. FUS = Transcranial ultrasound modeling step. Thermal = Thermal modeling step. }
  \label{figPerfomance1}
  \end{figure*}

\subsection{Results}  
Figure  \ref{figQceffect} compares the results of $T_S$ for each mapping method and selected values of correction coefficient $q_c$ to the experimental results of $T_E$. The best predictions with the Aubry, McDannold, Pichardo, and Webb-Marsac methods were found with correction values of $q_c$ of 0.4, 4, 3, and 3, respectively. While showing a decreasing trend with frequency, the figure also shows how literature values for $T_E$ are very dispersed for similar frequencies. The Webb-Marsac mapping method showed the most spread of predictions in $T_S$ to the SDR, capturing reported values of $T_E$ better. As shown in Fig. \ref{figDetailComparisonMap}, this result is more evident when plotting the individual results per SDR value. The Webb-Marsac mapping method was the only method that produced $T_S$ predictions proportional to SDR for frequencies larger than 0.4 MHz, suggesting it can better capture the effects of SDR in the total attenuation losses. The average distance (unitless) of $T_E$ to the predicted area $T_S$ values across all 5 SDR datasets for Aubry, McDannold, Pichardo, and Webb-Marsac methods were 0.041, 0.045, 0.43, and 0.021 respectively. Figure \ref{figDetailComparisonMap} also shows the values of $T_S$ obtained using a simplified mask with only T1w and T2w as input imaging, which showed a trend at the center of reported values of $T_E$.

\section{Performance study}
\subsection{Methods}
A numerical study was performed to analyze the computing performance of BabelBrain across different scenarios.
\subsubsection{Tests for transducer type, use of CT for mask generation, frequency, and refocus} 
Performance tests of processing wall-time were performed for all four transducers currently supported in BabelBrain. Testing was done with the dataset for SDR=0.55, same brain target, and transducer as section \ref{mappingsection}. For the simple transducer, tests were performed with frequencies between 0.2 and 1 MHz. For the other transducers, tests were done for their supported frequencies.
For the H317 transducer, tests also considered the refocusing step as described in subsection \ref {transcranial-modeling}. All tests were done at 6 PPW and included a T1w+T2w+CT-based mask as input imaging and T1w+T2w simplified mask for the case of a single transducer, with a domain size of 54$\times$54$\times$76 mm. Thermal simulations were performed for a sonication duration of 40 s, duty cycle of 30\% and cooling of 80 s. Simulations were done with a MacBook Pro M1 Max with 64 GB RAM. 

\subsubsection{Tests between different GPU systems} Processing wall-time was measured for different computer systems including: iMac Pro with 128 GB RAM, macOS 13.1, 3 GHz Xeon-W with 8 cores, with an internal Vega56 GPU (AMD) with 6 GB RAM, and an external W6800 Pro GPU with 32 GB RAM; MacBook Pro M1 Max with 64 GB RAM, macOS 13.1; Dell system with 128 GB RAM, Windows 11 and Linux (via WSL2), 4 GHz Xeon-W 2125 and CUDA 11.8. Tests were done with the dataset for SDR=0.41, with the same brain target and transducer as in section \ref{mappingsection}, a frequency of 600 kHz at 6 PPW, and a domain size of 54$\times$54$\times$76 mm. All systems used Anaconda Python manager with Python 3.10 (native for each CPU technology) and the same versions of supporting Python libraries (Numpy v1.23.5, SciPy v1.9.1, PySide 6.4.2, to mention a few) required to execute the software.
 
 \subsection{Results}
Figure \ref{figPerfomance1} shows the consolidated results of wall-time by type of transducer, refocusing, frequency, and type of mask. The most time-consuming step is the domain generation that took in average ($\pm$s.d.) 72($\pm$18)\%, followed by the transcranial modeling with 24($\pm$16)\% and thermal modeling with 4($\pm$5)\%. While some tasks in the domain generation are GPU-accelerated, such as voxelization and median filters, functions for bias correction, coregistration, and upscaling are not GPU-accelerated and have a noticeable impact on performance. Analysis of results of the single transducer as a function of frequency and use of CT indicated that the domain-generation step had a $O^3$ ($R^2>0.99$) proportionality in computing time with the frequency. In contrast, the transcranial modeling step had a proportionality of $O^2$ ($R^2>0.99$). The use of CT scans also had an important impact of 1 order of magnitude in the domain generation step, given the extra steps required in its processing. For frequencies of interest such as 500 kHz, the total processing time could vary from 55 s when using a simplified mask to 7.5 min when using a CT scan. 

\begin{figure}[!t]
  \centerline{\includegraphics[width=0.5\columnwidth]{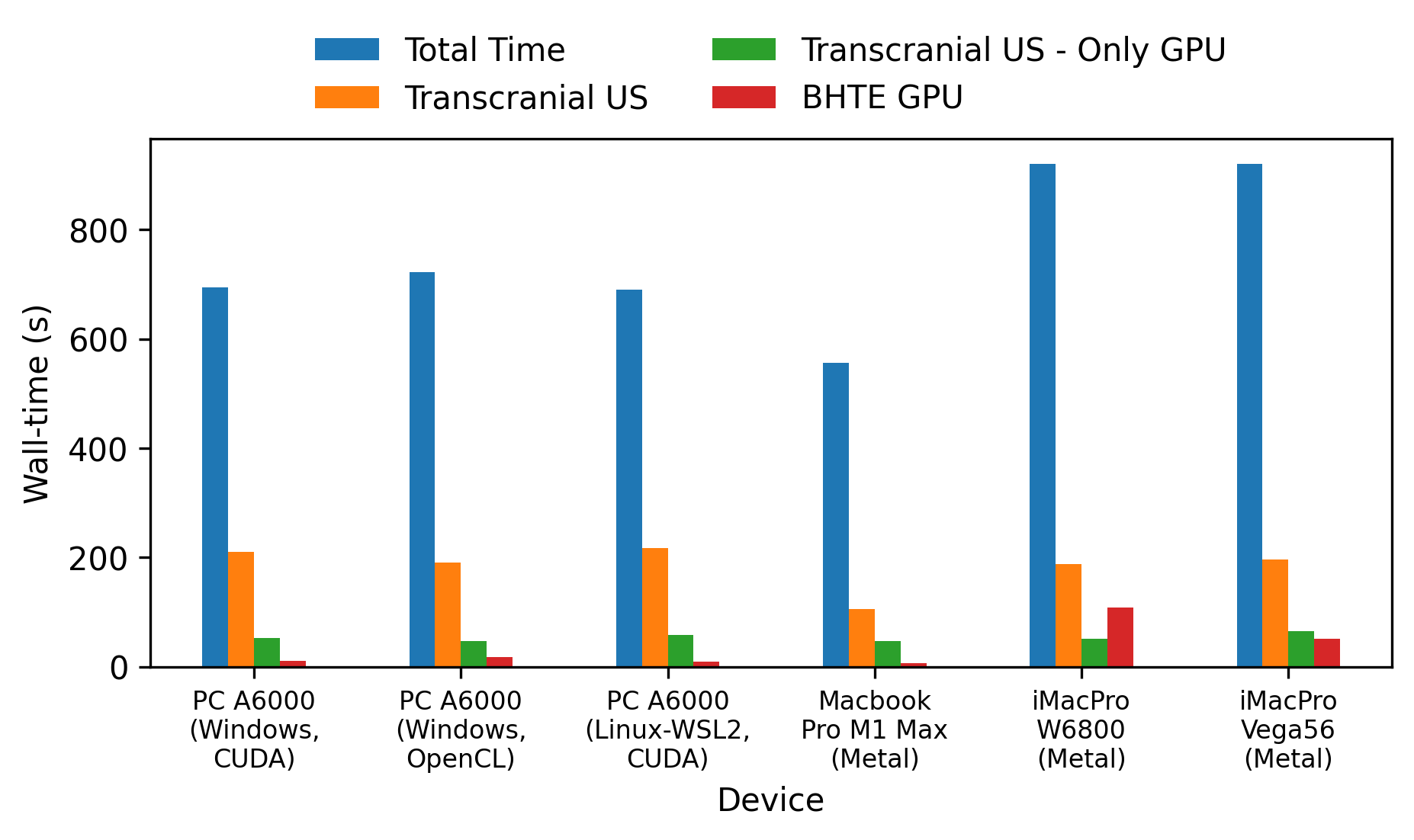}}
  \caption{Performance results with different GPU systems using identical simulation conditions for a single transducer (focal length=50mm, F\#=1, 500 kHz) for a subcortical target in the dataset for SDR=0.42.}
  \label{figPerfomance2}
  \end{figure}
  Figure \ref{figPerfomance2} shows the performance results across several GPU computing systems configurations under identical conditions, including a more detailed breakdown of the time used exclusively in GPU calculations. Results indicated that non-GPU-accelerated calculations dominated the wall-time processing. GPU transcranial and thermal modeling calculations represented a small fraction of 11($\pm$3)\% of the total computational time across all configurations. The M1 Max-based system showed the best performance with a total computing time of 556s and GPU-alone time of 55 s, showing a global 20\% better performance and 15\% GPU-alone better performance when compared to the Dell system with an A6000 GPU. For the latter, results indicated that there was just a marginal difference between Windows+CUDA, Windows+OpenCL, or Linux-WSL2+CUDA. 
Performance for the GPU-alone part of the transcranial modeling step showed similar results across all GPU devices, from the M1 Max device requiring 51 s to the A6000 (Linux-WSl2, CUDA) requiring 58 s. For BHTE GPU calculations, however, the AMD GPUs showed much worse performance, with 117 s and 159 s with the Vega56 and W6800 GPUs, respectively. This performance contrasts the 7 s and 9 s required with the M1 Max and A6000 processors, respectively.

\section{Effects of mask generation method }
\subsection{Methods}
A test was performed to study the effects of using a simplified mask using T1w+T2w alone compared to CT. The five datasets and targets used in section \ref{mappingsection} were tested with the CTX-500 device to evaluate how much correction needed to be done to ultrasound focusing. The transducer was assumed to be in contact with the skin. Electronic steering and mechanical adjustments were made in BabelBrain to ensure the focal spot was at the target. Measurements were done for focal zone dimensions at -6dB and peak pressure. Thermal simulations were performed for a sonication duration of 40 s, duty cycle of 30\%, and cooling of 80 s.

\subsection{Results}
\begin{figure}[!t]
  \centerline{\includegraphics[width=0.5\columnwidth]{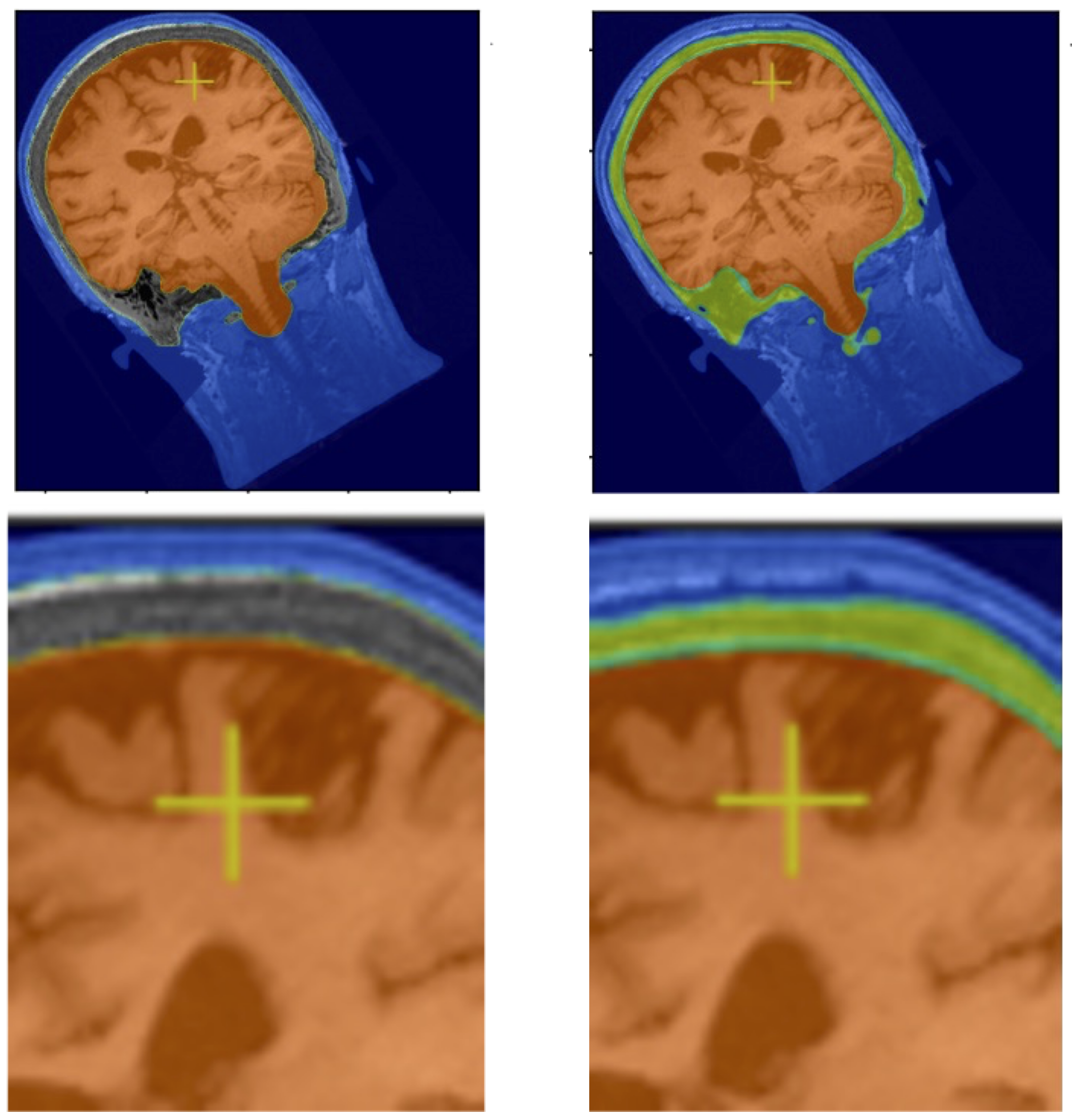}}
  \caption{Comparison of generated masks for the case of SDR=0.55. Left) Mask produced with CT scan where an overlay of different regions, T1w, and CT data are shown. Right) Simplified mask. In-plane view (top) along the trajectory and zoom (bottom) of regions close to the target ($+$). }
  \label{figMasks}
  \end{figure}
\begin{figure*}
  \centerline{\includegraphics[width=\columnwidth]{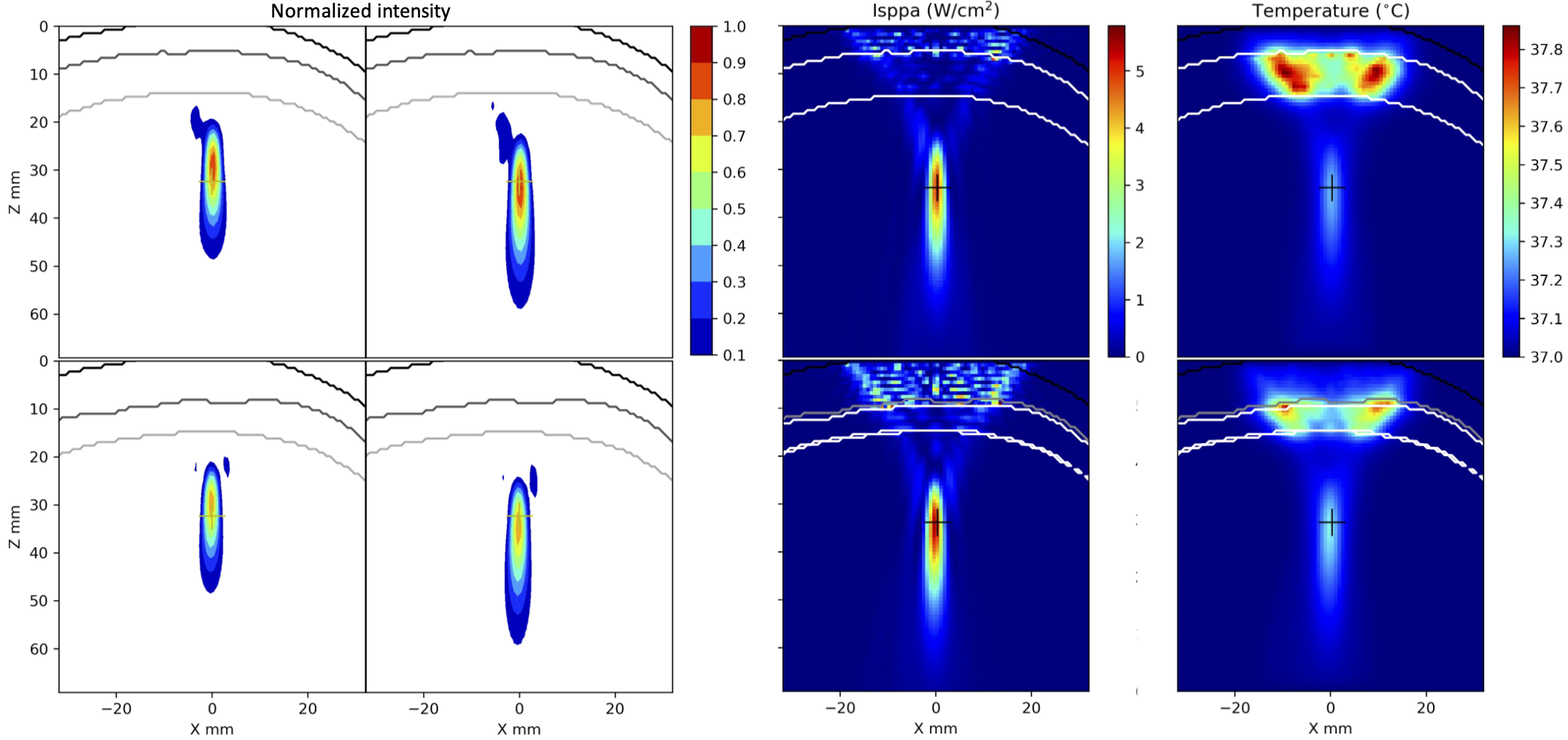}}
  \caption{$xz$ acoustic intensity and thermal maps at $y=0$ showing examples of corrections applied for dataset with SDR=0.55 with CT-based simulation (top row) and simplified mask (bottom row). Target indicated by $+$ mark. Normalized intensity as shown in step 2 of BabelBrain with steering calculated at the distance (34.2 mm) of the target from the skin (first column). Normalized intensity after applying a steering correction of 5 mm in depth (second column). Acoustic intensity (third column) and thermal maps (fourth column) at the end of the sonication (40s) for a selected $I_{\text{SPPA-brain}}=5$W/cm$^2$.  }
  \label{figCorrections}
  \end{figure*}

Figure \ref{figMasks} shows an example of the masks produced for the dataset for SDR=0.55. The simplified mask derived from `charm` tool output showed a slightly thinner skull region when compared to CT. The average difference in steering applied in the CTX-500 device to reach the desired target between the simple mask and CT was 1.00($\pm$1.3) mm. The average difference in mechanical repositioning in $x$ and $y$ directions was, respectively, 0.2($\pm$0.4) and 0.1($\pm$0.5) mm. These results indicated that the simple mask was adequate to estimate a correction level in steering and repositioning that was similar to when using CT.  However, the dimensions of ultrasound focus and peak pressure showed larger differences between simple mask and CT processing. Figure \ref{figCorrections} shows an example of acoustic intensity maps before and after steering and mechanical positioning corrections, along with thermal effects. Figure \ref{figCorrections} also shows that while the focus is on the intended position, the intensity distribution is different between simple mask and CT. Table \ref{tableDiff} shows a summary of differences in focus dimensions calculated at -6dB and peak pressure. Overall, the use of simple masks created larger focus volumes. Peak pressure was overestimated for SDR$\leq$0.42 and underestimated for SDR$\geq$0.55, which was consistent with results shown in  Fig. \ref{figDetailComparisonMap}.  

  \begin{table}
  \caption{Difference in \%  in focus length, width, and peak pressure between results with simplified mask and CT. Positive and negative values indicate, respectively, overestimation and underestimation when simplified mask relative to CT.}
  \begin{center}
  \begin{tabular}{lrrrr}
    \toprule
    {} &  Focus  &  Focus  &  Focus  &  Peak  \\
    SDR  &           length         &   width                 &   volume    &   pressure   \\
    \midrule
    0.31 &                4.7 &                3.6 &   4.2 &  24.4 \\
    0.42 &               11.1 &               -0.3 &  10.9 &   8.0 \\
    0.55 &                0.8 &                0.2 &   1.9 & -14.7 \\
    0.67 &               17.9 &                5.6 &  40.1 &  -9.3 \\
    0.79 &               -4.3 &                4.5 &  -5.0 & -10.6 \\
    \midrule
    \emph{mean ($\pm$s.d)}&  6($\pm$8)&  2.7($\pm$2) &  10.4($\pm$16) &-0.5($\pm$15)\\
    \bottomrule
    \end{tabular}
  \end{center}
  \label{tableDiff}
  \end{table}
$~$\par
$~$\par

\section{Discussion} 
\subsection{Selection of mapping method}
Mapping procedures of acoustic properties for transcranial ultrasound will undoubtedly remain an important research topic, especially for attenuation. Previous work demonstrated that existing mapping methods from HU to speed of sound offer a similar level of prediction in terms of phase aberration \cite{bancel_comparison_2021}. In contrast, studies for mapping of attenuation have shown significant differences \cite{mcdannold_elementwise_2019,webb_acoustic_2021}. In this study, the combination of the BabelBrain modeling procedure with the mapping methods proposed by Aubry, McDannold, Pichardo, and  Webb-Marsac were tested to estimate the pressure transmission of transcranial ultrasound.   
The results with the Webb-Marsac method showed a transmission proportional to the SDR (the lower the SDR, the higher the losses) and closer to the transmitted coefficients reported in experimental studies. However, correction factors $q_c$ between 3 and 4 had to be applied for the mapping methods by McDannold, Pichardo, and Webb, as they produced attenuation values that were initially too high for the viscoelastic solver. In opposition, the Aubry method produced values that were too low, requiring a correction factor of 0.4. These corrections highlight the challenges of applying a given mapping method and the importance of accounting for imaging resolution and material heterogeneity. 
 
 A recent study found that when using simple tissue masks with homogenized skin, bone, and brain tissue layers, multiple modeling methods, including the one used in this study, produced similar results\cite{aubry_benchmark_2022} without needing any additional corrections. 
 However, when using voxel-level properties in a viscoelastic solver, the $Q$ factor models exclusively dissipative effects and scattering effects are introduced by the heterogeneity of the material. 
 This translates to a scenario where the mapping of acoustic properties using voxel-level heterogeneity may be specific to the solver in use, which could potentially explain the significant differences between attenuation mapping approaches in previous studies \cite{webb_acoustic_2021}. Based on the findings of this study, the default mapping procedure for the BabelBrain application in its first public release (BabelBrain v0.2.5) is the Webb-Marsac approach with a correction factor of $q_c=3$. For the specific FDTD solver of the viscoelastic equation used in BabelBrain, the Webb-Marsac method captured reported values of experimental pressure transmission the best, and it translated into a pressure transmission that was, for the most, proportional to the SDR score, which is in agreement with the accumulated evidence of transcranial therapy with MRgFUS\cite{chang2016factors,boutet2019relevance}. Given that for surgery applications with MRgFUS, other factors such as skull geometry and position of large aperture transducers have an effect on sound transmission, it is important to emphasize the observations in this study are exclusively for much smaller apertures and in normal incidence conditions.

One limitation of the study's methodology is that the reported values of $T_E$ in the literature are not contextualized to specific values of SDR. While it can be reasonably argued that, given the number of reports considered, the collected data represented a sampling of different SDR conditions, more rigorous pressure transmission data collection with multiple skull samples associated with SDR values is necessary to provide a more definitive conclusion.

The use of simple masks and homogenized material properties (Table \ref{tableMaterial}) produced, on average,  similar results to CT-based processing but with important variations across the tested datasets (Table \ref{tableDiff}).   Findings in this study when using a simplified mask compared to CT showed similar deviation (from $-$2.1\% to +16.5\%) in focus volume to a similar study using two skull specimens\cite{muellerNumericalEvaluationSkull2017}, while the same study showed higher deviation (from +18.3\% to 58.7\%) in peak pressure. Also important to mention, the validation of masks with homogenized material was only conducted in the parietal bone region, in normal incidence conditions, and with a device with a small aperture of 5 cm. Nevertheless, the reported s.d. in Table \ref{tableDiff} and plots in Fig. \ref{figQceffect} can provide guidance on the expected incertitude during FUS-based neuromodulation experiments if using only simple masks derived from T1w+T2w alone.

The CT scan conditions of the subject datasets used in this study are well represented in the fitting tables in Webb's studies \cite{webb_measurements_2018,webb_acoustic_2021}. While future versions of BabelBrain (or user-modified versions) could easily add a selection of different CT conditions of energy level, vendor, and kernel, a numerical study as the one performed here would be required to validate the accuracy of simulations across different SDR conditions. The availability of the datasets used in this study in a public repository can facilitate this testing.

As indicated in the Methods, for this study, the mapping of shear properties derived from CT scans was disabled. While some manual adjustments could have been applied to correct the voxels producing invalid Poison ratio coefficients, it seemed a detriment to the rigorousness desired for the proposed simulation tool. A study must be performed to recalculate the relationship of shear speed of sound where the Webb-Marsac functions for longitudinal speed of sound are used. Based on the results in \cite{bancel_comparison_2021}, where several mapping methods for longitudinal speed of sound produced similar results for phase correction purposes, it is reasonable to anticipate an updated relationship for shear properties will not deviate significantly from what was published in \cite{pichardoViscoelasticModelPrediction2017b}. 
  
\subsection{Current limitations of ZTE-based processing} 
It is important to emphasize that, in the current release of BabelBrain, the ZTE-based processing should still be considered more experimental compared to true CT-based processing. Future studies are needed where transcranial modeling with ZTE-based processing is compared to findings with real CT scans. However, the work in \cite{miscouridouClassicalLearnedMR2022} demonstrated that a simple conversion based on normalization of the ZTE signal produced better transcranial ultrasound predictions when compared to machine learning (ML) training methods using T1w scans. While ML methods with ZTE offer superior synthetic CT map reconstruction\cite{miscouridouClassicalLearnedMR2022}, ML often requires training using site-specific MRI scans, and CT scans availability is not always ensured. A method based on normalizing the ZTE signal has the advantage of not requiring training. However, as ML methods continue to improve \cite{miscouridouClassicalLearnedMR2022, liuEvaluationSyntheticallyGenerated2022, kohAcousticSimulationTranscranial2022, yaakubPseudoCTsT1weightedMRI2023}, there is a potential of reducing the need for training data from CT scans for studies with limited access to them.

\subsection{Computing Performance}
The goal of developing BabelBrain was to make it easy to use and efficient on a variety of computer systems, including those from different manufacturers. The code for the CUDA, OpenCL, and Metal languages has been optimized multiple times since the release of BabelViscoFDTD in 2020. Generally, the transcranial modeling step with BabelViscoFDTD had similar performance across high-end GPU models. Excepting for AMD devices, BHTE calculations were the fastest to perform. The M1 Max system had the best overall performance, including in GPU execution. Its high-speed unified memory bank between the GPU and CPU can partly explain this, which is faster than the common DDR4 memory technology used in high-end GPU systems. ARM-based GPUs like the M1 Max are also smaller, more portable, and less expensive compared to high-end discrete GPUs, such as the A6000. For frequencies and devices commonly used in neurostimulation (500 kHz or less, transducers with apertures less than 10 cm), the total computing time is around 10 minutes or less. Additionally, the most time-consuming step of domain generation does not need to be repeated in many cases, and corrections made in BabelBrain for ultrasound steering or manual repositioning only require a small fraction of the computing time. 
Future developments will explore improving the performance of the domain generation step, particularly regarding tasks such as bias correction, upscaling, and coregistration. Also, refinements to the simplified mask generation will be explored to improve the quality of skull bone mask. 
 
\subsection{Other potential applications of BabelBrain }
Beyond the intended use for prospective planning of experiments in humans for neuromodulation applications with FUS, BabelBrain's use can be easily extended for the planning of cavitation-mediated FUS brain therapy, such as the opening of the brain-blood barrier, using microbubbles and low-intensity FUS \cite{mengSafetyEfficacyFocused2019}. From a planning design perspective, both applications are practically equivalent. It is also worth mentioning, that BabelBrain could be also adapted as part of studies for forward and inverse problems; for example, in the study full-wave inversion in ultrasound computed tomography\cite{guaschFullwaveformInversionImaging2020}, but special attention must be done to ensure BabelBrain is combined with other methods to avoid the so-called inverse crime \cite{wirginInverseCrime2004}.

While tools such as BabelBrain can assist in the tasks to perform planning of FUS procedures, it is important to mention that in the context of optical-based neuronavigated procedures improvements remain to be done to increase the precision of optical targeting. Previous reports \cite{xuCharacterizationTargetingAccuracy2022,chaplinAccuracyOpticallyTracked2019,wuEfficientBloodBrainBarrier2018} indicate mistargeting of optical-based systems of a couple of millimeters, which needs to be improved to achieve more adequate target engagement of FUS-based neuromodulation.
  
\section{Conclusions}
In this study, the open-source research software BabelBrain was presented, which was developed to plan focused ultrasound procedures in humans using neuronavigation systems. A numerical study was done with datasets covering different values of skull density ratio to select the mapping procedure of acoustic properties derived from computed tomography. A study of numerical performance was conducted to illustrate the processing time achievable in modern computer systems. Finally, a study was done to illustrate the effects of using simplified masks compared to computed tomography-derived processing. BabelBrain and the imaging data used to conduct these studies are fully open-accessible to help researchers in their studies of focused ultrasound-based neuromodulation and other low-intensity applications of transcranial FUS, such as the opening of the brain-blood barrier. 

\bibliographystyle{IEEEtran}
\bibliography{refs}

\section*{Acknowledgment}
This work was supported in part by a Discovery Grant from the Natural Sciences and Engineering Research Council of Canada and the INOVAIT program.

The author thanks Ali K. Zadeh from the University of Calgary, Ghazaleh Darmani and Jean-Francois Nankoo from Toronto Western Hospital, Roch M. Comeau, Alex Ciobanu, and Sean McBride from Rogue Research for their feedback on the first iterations of the software. The author also thanks Andrew Xie for his work during the summer 2022 improving BabelViscoFDTD.

\end{document}